%% file: huard_charm2012_proceedings.tex
\documentclass[12pt]{article}
\usepackage{graphicx}
\usepackage{amsmath}
\usepackage{amssymb}
\usepackage{subfigure}
\usepackage{multirow}
\usepackage{multicol}

\RequirePackage{xspace}
\usepackage{relsize}
\def\babar{\mbox{\slshape B\kern-0.1em{\smaller A}\kern-0.1em
    B\kern-0.1em{\smaller A\kern-0.2em R}}}

\newcommand\pubnumber{}
\newcommand\pubdate{\today}

\def\cincinnati{Department of Physics\\
University of Cincinnati, Cincinnati, Ohio 45219, USA}

\newcommand{\ts}{\textsuperscript}

\def\Title#1{\begin{center} {\Large #1 } \end{center}}
\def\Author#1{\begin{center}{ \sc #1} \end{center}}
\def\Address#1{\begin{center}{ \it #1} \end{center}}

\newcommand\pubblock{\rightline{\begin{tabular}{l} \pubnumber\\
         \pubdate  \end{tabular}}}
\newenvironment{Abstract}{\begin{quotation}  }{\end{quotation}}
\newenvironment{Presented}{\begin{quotation} \begin{center} 
             PRESENTED AT\end{center}\bigskip 
      \begin{center}\begin{large}}{\end{large}\end{center} \end{quotation}}

\input{econfmacros.tex}
\begin{document}
\begin{titlepage}
\pubblock

\vfill
\Title{Precision measurements of charm hadron properties at \babar\ }
\vfill
\Author{Zachary Huard\\ on behalf of the \babar\ collaboration}
\Address{\cincinnati}
\vfill
\begin{Abstract}
We present preliminary results on the natural line width, $\Gamma\left(D^{*+}\right)$, in the transition $D^{*}\left(2010\right)^{+} \rightarrow D^0 \pi^+$.  The $D^0$ is reconstructed in two decay channels: $D^0 \rightarrow K^-\pi^+$ and $D^0 \rightarrow K^-\pi^+\pi^-\pi^+$. We use Monte Carlo to simulate our reconstruction resolution in the difference of the reconstructed $D^{*+}$ and reconstructed $D^0$ masses.  We fit a relativistic P-wave Breit-Wigner form convolved with the measured resolution, together with a background function, to the mass difference distribution. For the decay mode $D^0\rightarrow K^-\pi^+$ we obtain $\Gamma =  83.5 \pm 1.7 \pm 1.2$ keV and $\Delta m = m_{D^{*\,+}} - m_{D^0} = 145425.5 \pm 0.6 \pm 2.6$ keV, where the quoted errors are statistical and systematic, respectively. For the $D^0\rightarrow K^-\pi^+\pi^-\pi^+$ mode, we do not quote a width measurement at this time, but measure $\Delta m = m_{D^{*\,+}} - m_{D^0} = 145426.5 \pm 0.4 \pm 2.6 $ keV.  Additionally, we present preliminary results from the decay mode $\Lambda_c \left(2880\right)^+ \rightarrow \Lambda_c^+\pi^+\pi^-$ for the mass, width and quasi-two-body branching ratios of the $\Lambda_c\left(2880\right)^+$ baryon.  We measure $m\left(\Lambda_c\left(2880\right)^+\right) - m\left(\Lambda_c\left(2286\right)^+\right) = 594.9 \pm 0.4 \pm 0.3$ MeV and $\Gamma\left(\Lambda_c\left(2880\right)^+\right) = 6.2 \pm 1.4 \pm 1.5$ MeV, where the errors are statistical and systematic, respectively.
\end{Abstract}
\vfill
\begin{Presented}
Charm 2012\\
5\ts{th} International Workshop on Charm Physics\\
Honolulu, Hawai'i, USA, May 14 - 17, 2012
\end{Presented}
\vfill
\end{titlepage}
\def\thefootnote{\fnsymbol{footnote}}
\setcounter{footnote}{0}
%%%%%%%%%%%%%%%%%%%%%%%%%%%%%%%%%%%%%%%%%%%%%%%%%%%%%%%%%%%%

\input{dstarwidth.tex}

\input{lambdac2880.tex}
\section{Summary}
Our preliminary results of the $D^{*}\left(2010\right)^{+}$ width and RBW pole are
\begin{align*}
\Gamma\left(K\pi\right) &= 83.5 \pm 1.7 \pm 1.2 {\rm{\,keV}}\\
\Delta m\left(K\pi\right) &= 145425.5 \pm 0.6 \pm 2.6 {\rm{\,keV}} \\
\Delta m\left(K3\pi\right) &= 145426.5 \pm 0.4 \pm 2.6 {\rm{\,keV}}
\end{align*}
\noindent At this time, we do not report the width measurement for the $K3\pi$ mode pending investigation of the bias in the validation fit. The measurements of $\Delta m$ for the two modes are in excellent agreement, and have small systematic uncertainties. The uncertainties reported here are conservative overestimates and we anticipate they may be reduced further. Using our result and updated width measurements for a few other states, we find large disagreements among the calculated values of the universal coupling constant using the ratios provided by Ref. \cite{PhysRevD.64.114004}.\par

Using data on the decay $\Lambda_c \left(2880\right)^+ \rightarrow \Lambda_c \pi^+\pi^-$, we find preliminary values for the mass difference $\Delta M = m\left(\Lambda_c\left(2880\right)^+\right) - m\left(\Lambda_c\left(2286\right)^+\right) = 594.9 \pm 0.4 \pm 0.3$ MeV and the width $\Gamma = 6.2 \pm 1.4 \pm 1.5$ MeV. These values are consistent with the previous \babar\ results from an analysis of the $D^0p$ mode \cite{PhysRevLett.98.012001}. We also report the first measurements of the branching ratios to the quasi-two-body $\Sigma_c \pi$ and non-resonant $\Lambda^+\pi^+\pi^-$ states.

%%%%%%%%%%%%%%%%%%%%%%%

\end{document}

%% file: econfmacros.tex
\def\beq{\begin{equation}}
\def\eeq#1{\label{#1}\end{equation}}
\def\eeqn{\end{equation}}

\def\beqa{\begin{eqnarray}}
\def\eeqa#1{\label{#1}\end{eqnarray}}
\def\eeqan{\end{eqnarray}}

\let\bar=\overbar

\def\Dslash{\not{\hbox{\kern-4pt $D$}}}
\def\dslash{\not{\hbox{\kern-2pt $\del$}}}

\def\msb{{\bar{\ssstyle M \kern -1pt S}}}

%% file: dstarwidth.tex
\section{$D^{*}\left(2010\right)^{+}$ line width}

The $D^{*+}$ line width, $\Gamma\left(D^{*+}\right)$, provides an experimental check of spectroscopic models, and is also related to the strong coupling $g_{D^* D \pi}$. In the heavy-quark limit this coupling can, in turn, be related to the universal coupling $g$ \cite{PhysRevC.83.025205}. There is no direct experimental window on the corresponding coupling in the $B$ system, $g_{B^*B\pi}$, since there is no phase space for the decay $B^* \rightarrow B\pi$.  However, the $D$ and $B$ systems can be related through $g$, and this opens an avenue to the calculation of $g_{B^*B\pi}$ \cite{PhysRevC.83.025205}.\par

We study $D^{*+}\rightarrow D^0 \pi^+$ decays for two $D^0$ decay modes, and extract values of the width and $\Delta m$, the mass difference between the reconstructed $D^{*+}$ and reconstructed $D^0$ \cite{ccimplied}.  The detector resolution for $\Delta m$ is approximately $300$ keV, which is larger than the previously measured value of $\Gamma\left(D^{*+}\right)$ which was only approximately $90$ keV.  Therefore, to extract $\Gamma\left(D^{*+}\right)$ we fit the $\Delta m$ signal with a relativistic Breit-Wigner (RBW), which is described in the next section, convolved with a resolution function taken from a {\tt{GEANT4}} Monte Carlo simulation (MC) of the detector response. The key to this analysis is that, while the core resolution is larger than the extracted width, the RBW tails allow us to use information far from the central signal region to define the width. The RBW distribution, shown in Eq. (\ref{eq:rbw}) below, contains the natural decay width $\Gamma_0$.\par

The best measurement of the width to date is $\Gamma = 96 \pm 4 \pm 22$ keV by the CLEO collaboration, where the errors are statistical and systematic respectively  \cite{Anastassov:2001cw}.  The largest source of systematic uncertainty for that analysis was the lack of statistical significance of the tests used to investigate systematic effects.  In this analysis, we have much larger sample sizes for all modes investigated, which allows us to make tight selections and to investigate sources of systematic uncertainty with high precision.\par
%%%%%%%%%%%%%%%%%%%%%%%%%%%%%%%%%%%%%%%%%%%%%%%%%%%%%%%%%%%%%%%%%%%%%
\subsection{Data Selection and Fit Procedure}

The measurements are based on an integrated luminosity of approximately $480$ fb$^{-1}$ recorded at or near the $\Upsilon\left(4S\right)$ resonance by the \babar\ detector at the asymmetric energy PEP-II $e^+e^-$ collider. The \babar\ detector is described in detail elsewhere~\cite{ref:babar}. We reconstruct the decay $D^{*+}\rightarrow D^0 \pi_s^+$ in two prompt charm channels for the $D^0$.  The pion from $D^{*+}$ decay is described as the ``slow pion'' because of the limited phase space available for this decay. The selection criteria were chosen to give a very good signal to background ratio in order to increase the sensitivity to the long RBW tails; we have not optimized them for statistical significance.\par

Our fitting procedure involves two steps.  The first step is to use a triple Gaussian and a threshold function to model the resolution due to track reconstruction by fitting the $\Delta m$ distribution using truth-matched Monte Carlo (MC) events.  This MC was generated with a width of $0.1$ keV, so that all of the observed spread is due to reconstruction effects. The threshold function includes a non-Gaussian component that describes events in which the $\pi_s$ decayed to a $\mu$ in flight and we use hits from both the $\pi$ and $\mu$ segments.  Accounting for these non-Gaussian events greatly improves the overall fit at low and high $\Delta m$ values relative to the signal peak.\par

We use a binned maximum likelihood method to fit the MC data with the function:
\begin{equation}
f_{th} T\left(\Delta m; p, \alpha \right) + f_1 G\left(\Delta m; \mu_1, \sigma_1\right) + f_2 G\left(\Delta m; \mu_2, \sigma_2\right) + \left(1 - f_1 - f_2\right) G\left(\Delta m; \mu_3, \sigma_3\right) 
\end{equation}
where the $G\left( x; \mu_i, \sigma_i \right)$ are Gaussian functions and $f_{th}, f_1, f_2$ are the fractions contributed by the threshold function, and the first and second Gaussian components, respectively.  The threshold function is defined as
\begin{equation}
T\left(\Delta m; p, \alpha\right) = \Delta m \, u^p\,  e^{\alpha \cdot u}
\end{equation}
where $u \equiv \left(\Delta m/\Delta m_0\right)^2 - 1$ and $\alpha$ is a slope parameter.\par

\begin{figure}[ht]
\centering
\subfigure[$D^0 \rightarrow K^-\pi^+$]{\includegraphics[scale=0.35]{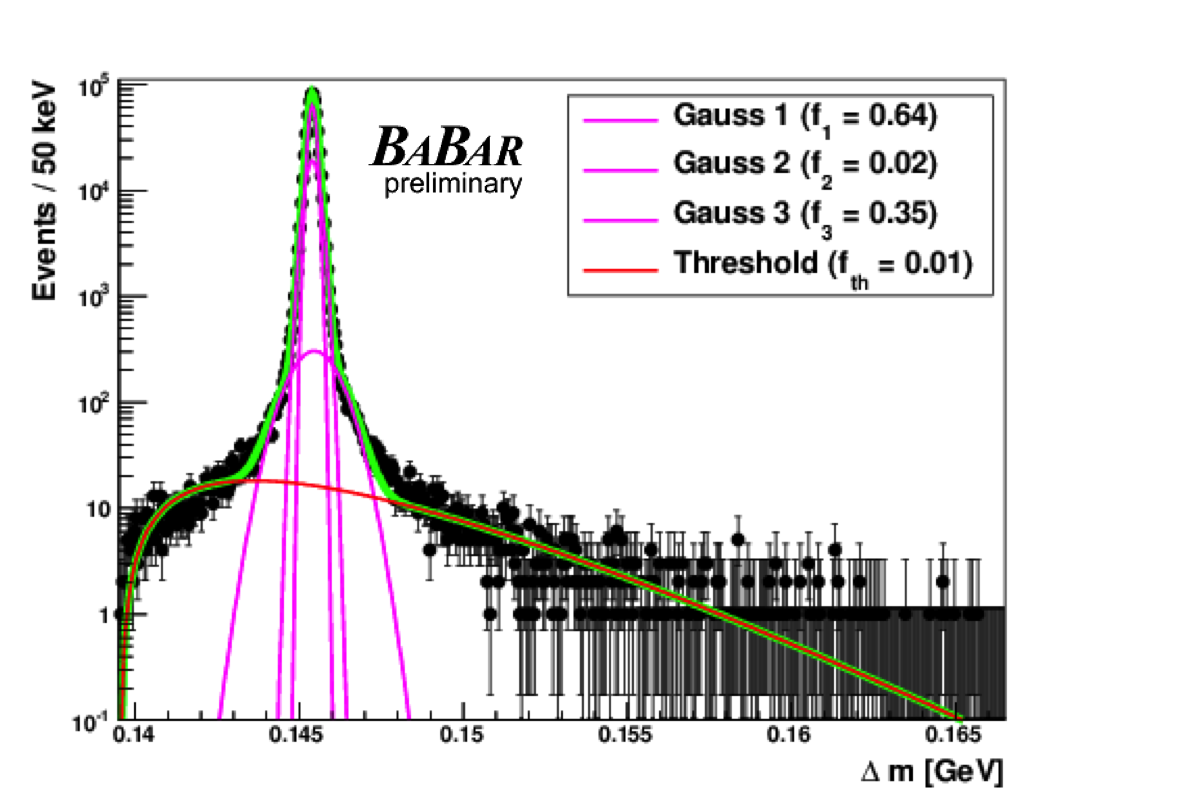}}
\subfigure[$D^0 \rightarrow K^-\pi^+\pi^-\pi^+$]{\includegraphics[scale=0.35]{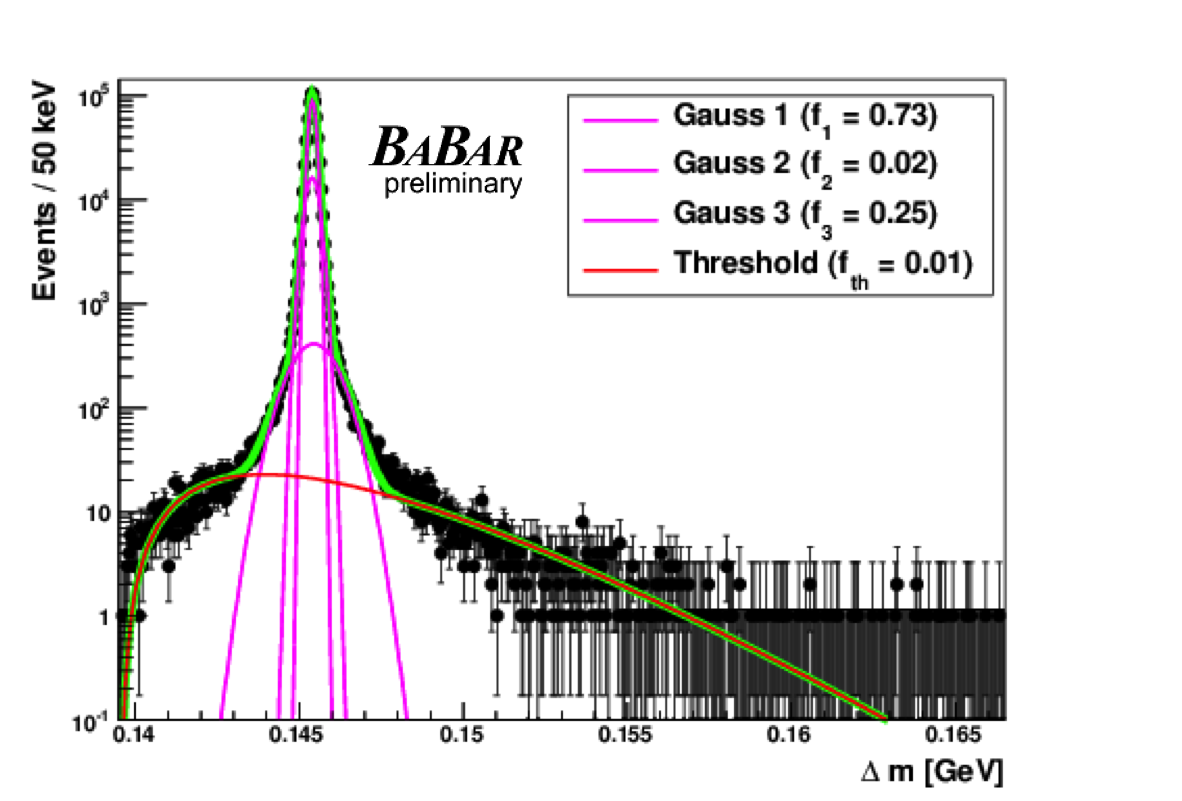}} \\
\caption{The fit to the MC resolution data  sample for both $D^0$ decay modes.  The interval size is $50$ keV, and the high mass tails are dominated by low statistics. Relative fractions for each PDF component are shown in the legend (rounded to the nearest 0.01).}
\label{fig:resfits}
\end{figure}

Figure \ref{fig:resfits} shows the individual resolution function fits for the two $D^0$ decay channels.  Each plot shows the total resolution probability density function (PDF) in green, the individual Gaussian contributions in magenta, and the threshold function in red.  The resolution functions should peak at the generated mass value; however, there is a slight bias toward high $\Delta m$ in the weighted mean of the three Gaussians (the threshold function accounts for less than 1\% of the signal and is excluded from this calculation).  We take this bias as an offset when extracting the position of the RBW pole from data.  The offset is only a few keV in size: 4.3 keV and 2.8 for $K\pi$ and $K3\pi$, respectively.\par

%%%%%%%%%%%%%%%

The second step is to take these resolution function parameters, convolve the Gaussian component with a relativistic Breit-Wigner (RBW) of the form given in Eq. (\ref{eq:rbw}), and fit the $\Delta m$ distribution in data. We fix the shape parameters and relative fractions from the resolution function in the fit to data.  There is additional flexibility in the fit to data as we allow a common offset, $\delta$, for the Gaussian mean values and a width scale factor, $\epsilon$, between MC and data. 

\begin{equation}
  \frac{d \Gamma(m)}{d m} =  \left(\frac{p}{p_0}\right)^{2\ell+1}  \left(\frac{1+r^2p_0^2}{1+r^2p^2}\right)
  \frac{\left(m_0 \Gamma_0\right)^2}  {\left(m_0^2  - m^2\right)^2  + \left(m_0 \Gamma _{\rm Total}(m) \right)^2}
  \label{eq:rbw}
\end{equation} 

\noindent The second factor is from the $D\pi$ form factor, $\mathcal{F}^{\ell}_{D\pi}$, represented by a Blatt-Weisskopf form factor with radius parameter $r$ \cite{blatt, PhysRevD.5.624}. We use the value $r = 1.6$ GeV$^{-1}$ from Ref. \cite{Schwartz:2002hh}. The specific value of the radius is unknown, but for the charm sector it is expected to be $\sim 1$ GeV$^{-1}$.  We vary this value as part of our investigation of systematic uncertainties.   Since the $D^{*+}$ is a vector particle we use $\ell = 1$.  $\Gamma _{\rm Total}(m)$ is defined by Eq. (\ref{eq:totalwidth}) and, subsequently, by Eq. (\ref{eq:partialwidth}).

\begin{equation}
  \Gamma_{\rm Total}(m) \approx \Gamma_{D^{*}D \pi}(m)
  \label{eq:totalwidth}
\end{equation}

\begin{equation}
  \Gamma_{D^*D \pi}(m) = \Gamma_0
  \left(\frac{\mathcal{F}^{\ell}_{D\pi}(p)}{\mathcal{F}^{\ell}_{D\pi}(p_0)}\right)^2\left(\frac{p}{p_0}\right)^{2\ell+1}\left(\frac{m_0}{m}\right)
  \label{eq:partialwidth}
\end{equation}

\begin{figure}[ht]
\centering
\subfigure[$D^0 \rightarrow K^-\pi^+$]{\includegraphics[scale=0.5]{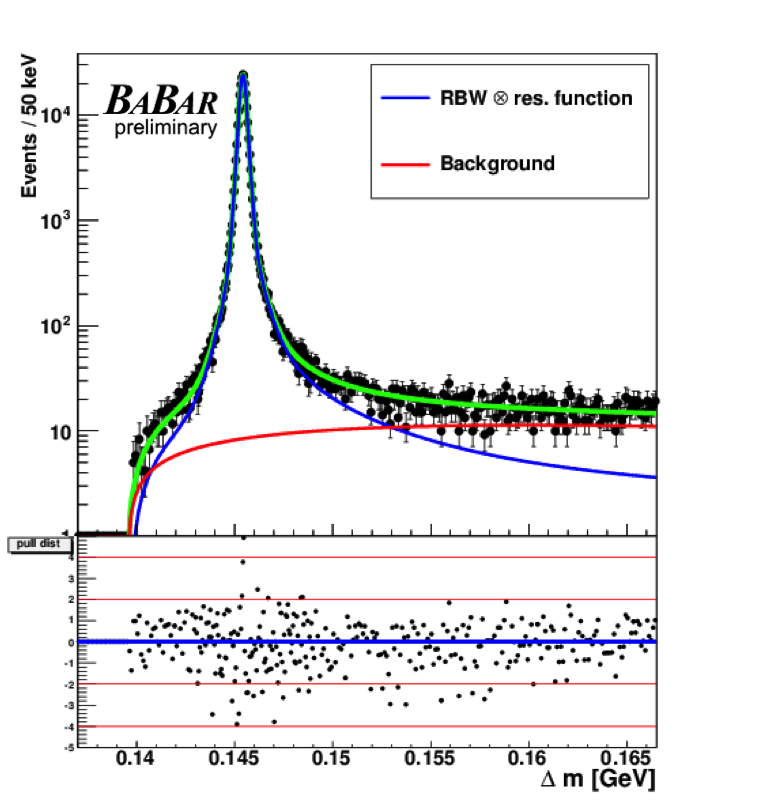}}
\subfigure[$D^0 \rightarrow K^-\pi^+\pi^-\pi^+$]{\includegraphics[scale=0.5]{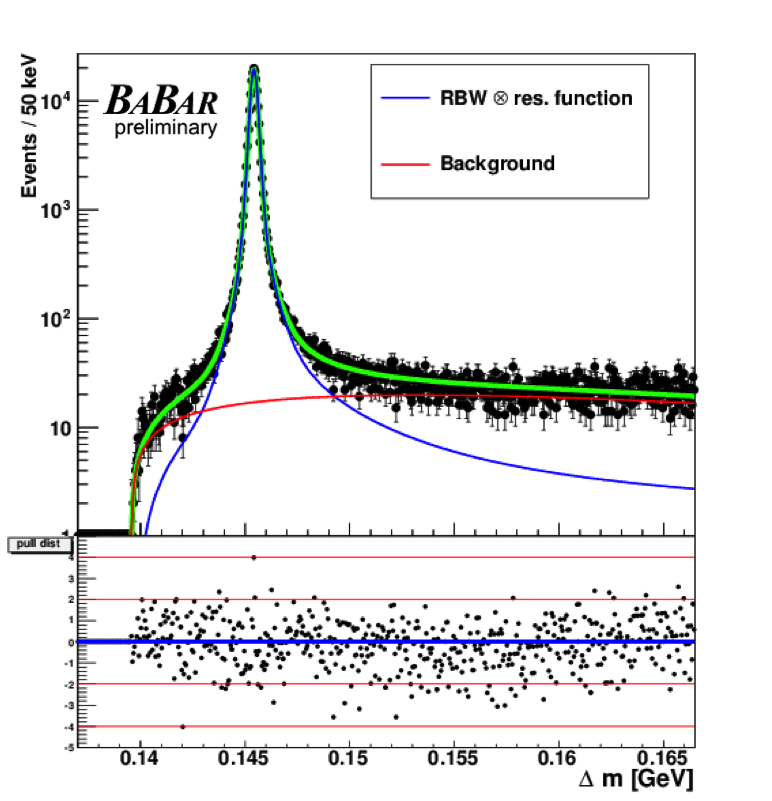}} \\
\caption[Data fits to both modes]{The fit to data for both $D^0$ decay modes (shown with logarithmic y axis scale). The total PDF is shown in green, the signal component in blue, and the background in red. The $\Delta m$ interval size is 50 keV. The normalized residuals, shown below each figure, are defined as $\left(N_{observed} - N_{predicted}\right)/\sqrt{N_{observed}}$.}
\label{fig:rdfits}
\end{figure}

\noindent Here, as in most applications, we approximate $  \Gamma_{\rm Total}(m) \approx \Gamma_{D^*D \pi}(m)$, by ignoring the electromagnetic contribution from $D^{*+}\rightarrow D^+ \gamma$. This approximation has a negligible effect on the extracted values.  We normalize the resulting PDF such that, at the pole $m=m_0$, Eq. (\ref{eq:rbw}) is equal to $1.0$.  Overall scale factors (such as factors of $\pi$, etc.) are swept into the normalization. Figure \ref{fig:rdfits} shows the fits to data for both channels.  The normalized residuals show very good agreement between the data and our model.\par

%%%%%%%%%%%%%%%%%%%%%%%%%%%%%%%%%%%%%%%%%%%%%%%%%%%%%%%%

\subsection{Systematic uncertainties}

The systematic errors are quantified independently for the $D^0\rightarrow K\pi$ and $D^0\rightarrow K3\pi$ modes, and are summarized in Table \ref{table:sys}.\par

For our studies of $\Gamma$ and the RBW pole as a function of $p_{lab}$, azimuthal angle $\phi$, and reconstructed $D^0$ mass, we split the data into disjoint subsets. These particular variables were chosen for careful study based in on dependency issues observed in previous \babar\ analyses. We ascribe an error for the variation observed in each variable using a procedure similar to the scale factor method used by the PDG (see their discussion of unconstrained averaging \cite{ref:pdg2012}). The width value does not display an azimuthal dependence; however, the RBW pole does. We continue to investigate the origin of this dependence, and temporarily take half of the variation as the error. Looking in slices of the reconstructed $D^0$ mass and $p_{lab}$ we estimate small systematic errors on $\Gamma$ and the RBW pole.\par

Our choice of fit range in the $\Delta m$ distribution is arbitrary, and so we study the effect of systematically varying the endpoint.  The choice of radius in the Blatt-Weisskopf form factor was tested by reducing the description of the $D^{*+}$ to a point particle.  We also estimate the systematic uncertainties which result from the modeling of the detector material and from the measurement of the magnetic field. The resolution function parameters extracted from the MC fit are also varied.  Since we use a binned maximum likelihood fit, we also double the number of intervals used in our fit as a check of the sensitivity to the nominal interval size of 50 keV. \par

For our validation fits, we generate signal MC samples with values for $\Gamma\left(D^{*+}\right)$ and $m\left(D^{*+}\right) - m\left(D^{0}\right)$ close to the current world averages \cite{ref:pdg2012}. Each signal MC sample is paired with a background cocktail in the same ratio as in the corresponding fits to data.  The validation data samples are approximately 5 times larger than the data. For each $D^0$ decay mode, the validation fits describe the data very well. In the fit to $K3\pi$, however, we do observe a bias in the extracted width. To be conservative we take the full difference between the fitted and generated value of the width, and assign a $5.7$ keV error. At present, we are investigating the source of this bias, which we hope to minimize.\par

\begin{table}[h!]
\begin{center}
\begin{tabular}{|c|c|c||c|c|}
\hline
\multirow{2}{*}{Source} & \multicolumn{2}{|c||}{$\sigma_{sys} \left(\Gamma\right)$ [keV]} & \multicolumn{2}{|c|}{$\sigma_{sys} \left({\rm{RBW\, pole}}\right)$ [keV]}\\ \cline{2-5}
 & $K\pi$ & $K3\pi$ & $K\pi$ & $K3\pi$ \\ \hline \hline
Detector material and magnetic field & 0.3 & 1.3 & 0.7 & 0.75 \\ \hline
$p_{lab}$ dependence & 0.57 & 0.19 & 0.2 & 0.2 \\ \hline
$m\left(D^0_{reco}\right)$ dependence & 0.0 & 1.12 & 0.01 & 0.0 \\ \hline
Azimuthal dependence & 0.45 & 0.65 & 2.5 & 2.5 \\ \hline
$\Delta m$ fit range & 0.8 & 0.3 & 0.1 & 0.0  \\ \hline
Resolution shape parameters & 0.41 & 0.3 & 0.17 & 0.14  \\ \hline
Blatt-Weisskopf radius & 0.04 & 0.03 & 0.0 & 0.0 \\ \hline
$\Delta m$ interval size & 0.1 & 0.1& 0.1&0.1 \\ \hline
MC validation & 0.0 & 5.7 & 0.0 & 0.0 \\ \hline \hline
Total & 1.2 & 6.0 & 2.6 & 2.6 \\ \hline
\end{tabular}
\end{center}
\caption{Summary of systematic errors. Total error is the sum in quadrature of each column.}
\label{table:sys}
\end{table}

%%%%%%%%%%%%%%%%%%%%%%%%%%%%%%%%%%%%%%%%%%%%%%%%%%%%%%%%%

\subsection{Results}
The preliminary results of the fits are
\begin{align*}
\Gamma\left(K\pi\right) &= 83.5 \pm 1.7 \pm 1.2 {\rm{\,keV}}\\
\Delta m\left(K\pi\right) &= 145425.5 \pm 0.6 \pm 2.6 {\rm{\,keV}} \\
\Delta m\left(K3\pi\right) &= 145426.5 \pm 0.4 \pm 2.6 {\rm{\,keV}}
\end{align*}

\noindent At this time, we do not report the width measurement for the $K3\pi$ mode pending investigation of the bias in the validation fit. The measurements of $\Delta m$ for the two modes are in excellent agreement, and have small systematic uncertainties. The uncertainties reported here are conservative overestimates and we anticipate that they may be reduced further.\par

The $D^{*+}$ line width can be related to the effective coupling of a pseudoscalar to a quark.  The total line width includes contributions from the $D^{*+}\rightarrow D^0 \pi^+$, $D^{*+}\rightarrow D^+ \pi^0$, and $D^{*+}\rightarrow D^+ \gamma$ decay channels. The strong couplings $g_{D^*D^0\pi^+}$ and $g_{D^*D^+\pi^0}$ are related through isospin conservation. Using chiral perturbation theory and the heavy-quark limit on the $c$ quark, this $D^{*+}$ strong coupling can be related to a universal coupling constant, $g$.  Di Pierro and Eichten obtain predictions for the value of this universal coupling constant in a variety of decay modes in terms of a ratio, $R$, which involves the width of the initial state \cite{PhysRevD.64.114004}. The coupling constant should then take the same value for the three $D$ decay channels listed in Table \ref{table:eichten}, which shows the values of the ratio $R$ extracted from the model and the experimental values for $\Gamma$, as they were in 2001. The values of $g$ obtained were in reasonable agreement, and indeed, at the time of publication, $g$ was consistent among all of the modes in Ref. \cite{PhysRevD.64.114004}.  In 2010, \babar\ published much more precise results on the $D_1\left(2460\right)^0$ and $D_2^*\left(2460\right)^0$ \cite{PhysRevD.82.111101}.  Using those results, our preliminary $\Gamma\left(D^{*+}\right)$ measurement, and the ratios from Table \ref{table:eichten} we can calculate new values for the coupling constant.  Table \ref{table:zach_eichten} shows the updated results. We estimate the error on the coupling constant value assuming $\sigma_\Gamma/\Gamma \ll 1$. The updated widths result in large disagreements among the calculated values of the universal coupling constant.\par

\begin{table}
\begin{center}
\begin{tabular}{|c|c|c|c|}
\hline
state & width ($\Gamma$) & $R=\Gamma/g^2$ (model) & g \\ \hline
$D^{*}\left(2010\right)^{+}$ & $96 \pm 4 \pm 22$ keV & $143$ keV & $0.82 \pm 0.09$ \\ \hline
$D_{1}\left(2420\right)^{0}$ & $18.9^{+4.6}_{-3.5}$ MeV & $16$ MeV & $1.09 ^{+0.12}_{-0.11}$ \\ \hline
$D_{2}^{*}\left(2460\right)^{0}$ &$23 \pm 5$ MeV & $38$ MeV & $0.77 \pm 0.08$ \\ \hline
\end{tabular}
\caption{Selected rows from Table 11 of Ref. \cite{PhysRevD.64.114004}.  State names correspond to the current PDG listings.  The third column is the ratio, $R$, extracted from the model in Ref. \cite{PhysRevD.64.114004}. The values of $g$ were obtained from the data available in 2001.}
\label{table:eichten}
\end{center}
\end{table}

\begin{table}
\begin{center}
\begin{tabular}{|c|c|c|c|}
\hline
state & width ($\Gamma$) & $R=\Gamma/g^2$ (model) & g \\ \hline
$D^{*}\left(2010\right)^{+}$ & $83.3 \pm 1.3 \pm 1.4$ keV & $143$ keV & $0.76 \pm 0.01$ \\ \hline
$D_{1}\left(2420\right)^{0}$ & $31.4 \pm 0.5 \pm 1.3$ MeV & $16$ MeV & $1.40 \pm 0.03$ \\ \hline
$D_{2}^{*}\left(2460\right)^{0}$ &$50.5 \pm 0.6 \pm 0.7$ MeV & $38$ MeV & $1.15 \pm 0.01$ \\ \hline
\end{tabular}
\caption{Updated coupling constant calculations using the latest width measurements. The values of $R$ are as in Table \ref{table:eichten}.  There is significant disagreement among the new values.}
\label{table:zach_eichten}
\end{center}
\end{table}

%% file: lambdac2880.tex
\section{Measurements of $\Lambda_c\left(2880\right)^+\rightarrow \Lambda_c^+ \pi^+\pi^-$}

A large number of charm baryons have been observed, but certain parameter values and decay characteristics still are not well-known \cite{ref:pdg2012}. In this analysis, we focus on the $\Lambda_c\left(2880\right)^+$ baryon, measure its mass and width, and search for resonant structures in its decay to the $\Lambda_c\left(2286\right)^+ \pi^+ \pi^-$ final state. The search yields the first measurements of quasi-two-body branching ratios of the decays $\Lambda_c\left(2880\right)^+\rightarrow \Sigma_c^0 \pi^+$ and $\Lambda_c\left(2880\right)^+\rightarrow \Sigma_c^{++} \pi^-$ where we reconstruct $\Sigma_c\left(2455\right)\rightarrow \Lambda_c\left(2286\right)\pi$ and $\Sigma_c\left(2520\right)\rightarrow \Lambda_c\left(2286\right)\pi$. Isospin symmetry demands equal branching fractions into the neutral and doubly charged modes.

\subsection{Data Selection and Fit Procedure}

The measurements are based on an integrated luminosity of approximately $316$ fb$^{-1}$ recorded at or near the $\Upsilon\left(4S\right)$ resonance by the \babar\ detector at the asymmetric-energy PEP-II $e^+e^-$ collider. The \babar\ detector is described in detail elsewhere \cite{ref:babar}. The selection criteria are optimized using the $S/\sqrt{S+B}$ metric on a 21.8 fb$^{-1}$ sample of $\Lambda_c^+\rightarrow p K^-\pi^+$ decays. This sample is excluded from the final analysis. To reduce B backgrounds we choose $x_p > 0.4$ where $x_p = p^*/p^*_{max} = p^*/\sqrt{\left(s/4\right) - m^2_{\Lambda_c}}$. Here, $p^*$ is the momentum in the center-of-mass (CM) frame, and $s$ is the CM energy squared.\par

\begin{figure}[h]
   \begin{center}
   \includegraphics[scale=0.4]{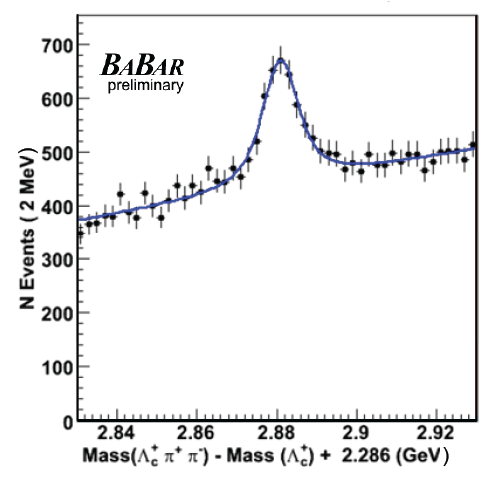} 
   \end{center}
   \caption{$\Delta M$ distribution for the region $x_p > 0.8$. The fit result is shown by the blue curve.}
   \label{fig:lambdam}
\end{figure}

To extract the mass and width, we fit the $\Delta M$ ($\equiv m\left(\Lambda_c^+\pi^+\pi^-\right) - m\left(\Lambda_c^+\right)$) distribution with a RBW convolved with a Gaussian resolution function and a polynomial background function. The Gaussian width is fixed to the detector resolution determined from MC. Figure \ref{fig:lambdam} shows the fit result for the region $x_p >0.8$, which has the highest signal-to-background ratio. The preliminary results are $\Delta M = 594.9 \pm 0.4 \pm 0.3$ MeV and $\Gamma = 6.2 \pm 1.4 \pm 1.5$ MeV where the errors are statistical and systematic, respectively. The systematic error is the sum of the uncertainties associated with the solenoidal component of the magnetic field, the tracking material, the background model, the signal model, the $\Delta M$ fit range, the $x_p$ momentum selection, and the detector resolution.\par

\begin{figure}[h]
   \begin{center}
   \includegraphics[scale=0.65]{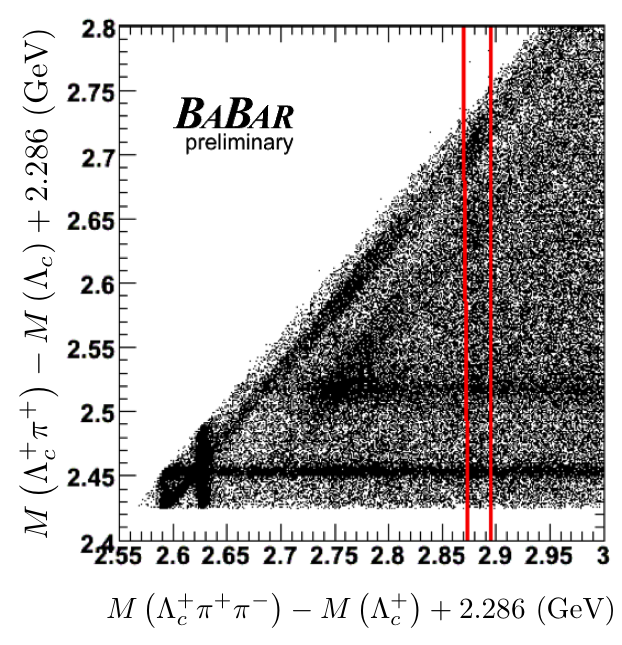} 
   \end{center}
   \caption{Mass spectrum of $\Lambda_c^+ \pi^+\pi^-$ invariant mass. Vertical bands correspond to $\Lambda_c\left(2593\right)^+$ and $\Lambda_c\left(2625\right)^+$ are clearly seen in the data. There is an enhancement in the $\Lambda_c\left(2765\right)^+$ region.  The red vertical lines enclose the relevant $\Lambda_c\left(2880\right)^+$ range. Horizontal bands correspond to $\Sigma_c\left(2455\right)^{++}$ and $\Sigma_c\left(2520\right)^{++}$ decays. The small Q value of the $\Sigma_c$ decays create a kinematic correlation of the $\Lambda_c$ and $\pi$, which creates the diagonal bands corresponding to $\Sigma_c\left(2455\right)^{0}$ and $\Sigma_c\left(2520\right)^{0}$ decays.}
   \label{fig:lambdascatter}
\end{figure}

Figure \ref{fig:lambdascatter} shows a number of resonant substructures, which are described in the caption. There are no visible overlaps in the $\Lambda_c\left(2880\right)^+$ signal region indicated by the red lines, and so we ignore interference effects when we calculate the related quasi-two-body branching ratios. We fit the $\Lambda_c\pi$ projections with a Gaussian for the $\Sigma_c$ signal and a polynomial background. Then we obtain the event yield by performing a background subtraction using the sidebands defined for each decay mode ($10-20$ MeV above and below the signal region). The raw yield is adjusted by an efficiency factor determined from MC to produce the corrected yield. Branching ratios are calculated according to Eq. (\ref{eq:lambf}). The non-resonant (NR) yield is the sum of the resonant yields subtracted from the total yield.\par

\begin{equation}
\frac{\mathcal{B}\left(\Lambda_c\left(2880\right)^+\rightarrow \Sigma_c\pi\right)}{\mathcal{B}\left(\Lambda_c\left(2880\right)^+\xrightarrow{all} \Lambda_c^+\pi^+\pi^-\right)}
= \frac{Y\left(\Lambda_c\left(2880\right)^+\rightarrow \Sigma_c\pi\right)}{Y\left(\Lambda_c\left(2880\right)^+\rightarrow \Sigma_c^+\pi^+\pi^-\right)} \times 
\frac{\epsilon\left(\Lambda_c\left(2880\right)^+\rightarrow \Sigma_c^+\pi^+\pi^-\right)}{\epsilon\left(\Lambda_c\left(2880\right)^+\rightarrow \Sigma_c\pi\right)}
\label{eq:lambf}
\end{equation}

Assuming no interference effects, and that $BF\left(\Sigma_c\rightarrow\Lambda_c\pi\right) = 100\%$, the preliminary branching ratio values obtained are:
\begin{align*}
\mathcal{B}\left(\Sigma_c\left(2455\right)^0\pi^+\right)/\mathcal{B}\left({\rm{all\,}} \Lambda_c^+\pi^+\pi^-\right) &= 0.13 \pm 0.02 \pm ^{+0.02}_{-0.01}\\
\mathcal{B}\left(\Sigma_c\left(2520\right)^0\pi^+\right)/\mathcal{B}\left({\rm{all\,}} \Lambda_c^+\pi^+\pi^-\right) &= 0.15 \pm 0.03 \pm ^{+0.02}_{-0.01}\\
\mathcal{B}\left(\Sigma_c\left(2455\right)^{++}\pi^-\right)/\mathcal{B}\left({\rm{all\,}} \Lambda_c^+\pi^+\pi^-\right) &= 0.2 \pm 0.03 \pm ^{+0.03}_{-0.01}\\
\mathcal{B}\left(\Sigma_c\left(2520\right)^{++}\pi^-\right)/\mathcal{B}\left({\rm{all\,}} \Lambda_c^+\pi^+\pi^-\right) &= 0.11 \pm 0.05 \pm ^{+0.02}_{-0.00}\\
\mathcal{B}\left({\rm{NR\,}}\Lambda_c^+\pi^+\pi^-\right)/\mathcal{B}\left({\rm{all\,}} \Lambda_c^+\pi^+\pi^-\right) &= 0.41 \pm 0.10 \pm ^{+0.08}_{-0.05}\\
\end{align*}

%% file: huard_charm2012_proceedings.bbl
\begin{thebibliography}{99}
\bibitem{ccimplied}
The use of charge conjugate reactions is implied throughout this paper.
\end{thebibliography}
